# Multimedia Traffic Routing in Multilayer WDM Networks

Angela Amphawan[*], Mohd Amirol Md Khair, Hassanuddin Hasan

InterNetWorks Research Laboratory, School of Computing, Universiti Utara Malaysia

*angela@uum.edu.my

**Abstract**

The advent of real-time multimedia services over the Internet has stimulated new technologies for expanding the information carrying capacity of optical network backbones. Multilayer wavelength division multiplexing (WDM) packet switching is an emerging technology for increasing the bandwidth of optical networks. Two algorithms for the routing of the multimedia traffic flows were applied: (i) Capacitated Shortest Path First (CSPF) routing, which minimizes the distance of each flow linking the given source and destination nodes and satisfying capacity constraints; and (ii) Flow Deviation Algorithm (FDA) routing, which minimizes the network-wide average packet delay.

## 1. Introduction

Wavelength division multiplexing (WDM) is an emerging technology for increasing the bandwidth of optical networks [1-4]. In multilayer WDM networks, traffic is carried over optical fiber connections which occupy a wavelength in each traversed fiber and terminates at an optical-to-electrical receiver at the destination node [5-10]. The connections are optically switched at the intermediate nodes and routing and wavelength assignment mechanisms are drawn on for determining the sequence of optical fibers traversed. The advent of real-time multimedia services over the Internet has stimulated new technologies for achieving the high level of Quality of Service (QoS) guarantee for sensitive multimedia traffic and for expanding the capacity of optical network backbones [11-22]. Various routing algorithms for reducing packet delays and alleviating network congestions for multimedia traffic have been developed [23-26]. The traffic flows in multilayer WDM networks take the form of multiprotocol label switching (MPLS) packets which are processed by electronic switching equipment. The electronic switching equipment is connected by optical connections. The set of optical connections established creates a virtual topology of optical connections, which is the topology governing the electronic equipment [27]. In virtual topology design, the set of optical connections required to transmit a given set of electronic traffic demands or electronic traffic flows are determined. With integrated optics, the implementation of active network components such as switches and multiplexers is viable. Packet switching features delivery of variable-bit-rate data streams through a sequence of packets over a shared network. When traversing switches, routers, network and other network nodes, packets are buffered and queued, resulting in variable delays and throughput depending on the multimedia traffic load in the network. In the top layer, the traffic flows are routed based on the virtual topology. In the lower layer, each optical connection in the virtual topology is routed over the physical topology and assigned a wavelength. In this paper, multimedia traffic will be introduced in the multilayer WDM network. Two algorithms for the routing of the multimedia traffic flows will be explored for minimizing the network-wide average packet delay, namely Capacitated Shortest Path First (CSPF) routing and Flow Deviation Algorithm (FDA) routing.

**2. Research Methodology**

A multilayer WDM network was simulated using MatPlanWDM [13]. The input parameters for the multilayer WDM network are: (1) the network topology, including the coordinates measured in kilometres over a Euclidean plane, node population, node type, number of nodes, number of time zone of each zone and the name of each node  (2) the multimedia traffic matrix, (3) the maximum number of transmitters and receivers in each node, (4) the maximum number of wavelengths in each link.  The two physical topologies used and their corresponding virtual topologies are shown in Fig. 1 and Fig. 2.   The total number of nodes is ten.   The number of fiber links is 34.

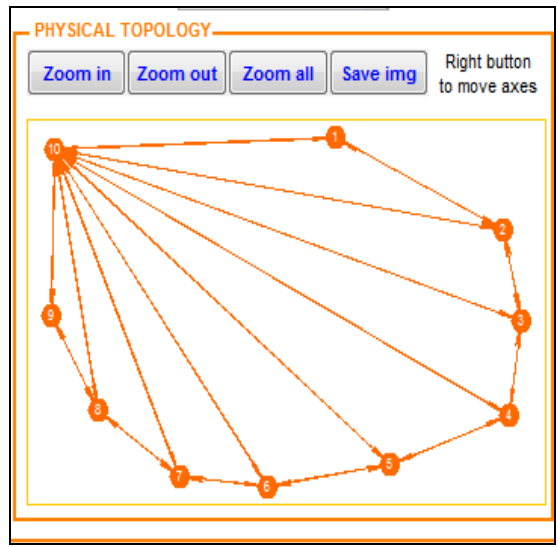

Figure 1.  Physical topology for Topology 1

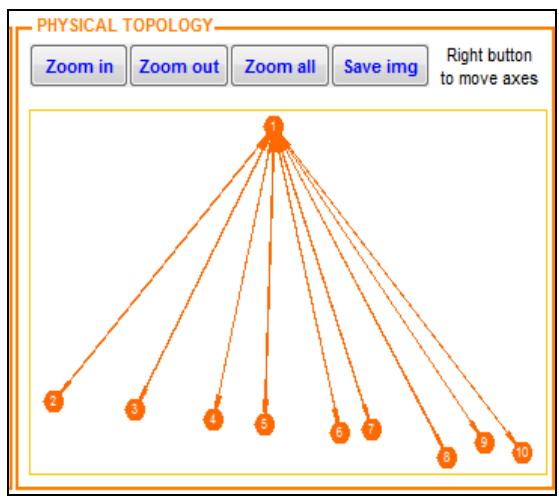

Figure 2. Physical topology for Topology 2

Audio traffic was generated using Markovian models. On-off models were applied for unitary flows while MMPP-N models were used for aggregated flows. Video traffic was generated using the "M/G/∞ input process" for unitary and aggregated flows.    The multimedia traffic matrices used for Topology 1 and Topology 2 are shown in Fig. 3 and Fig. 4 respectively.

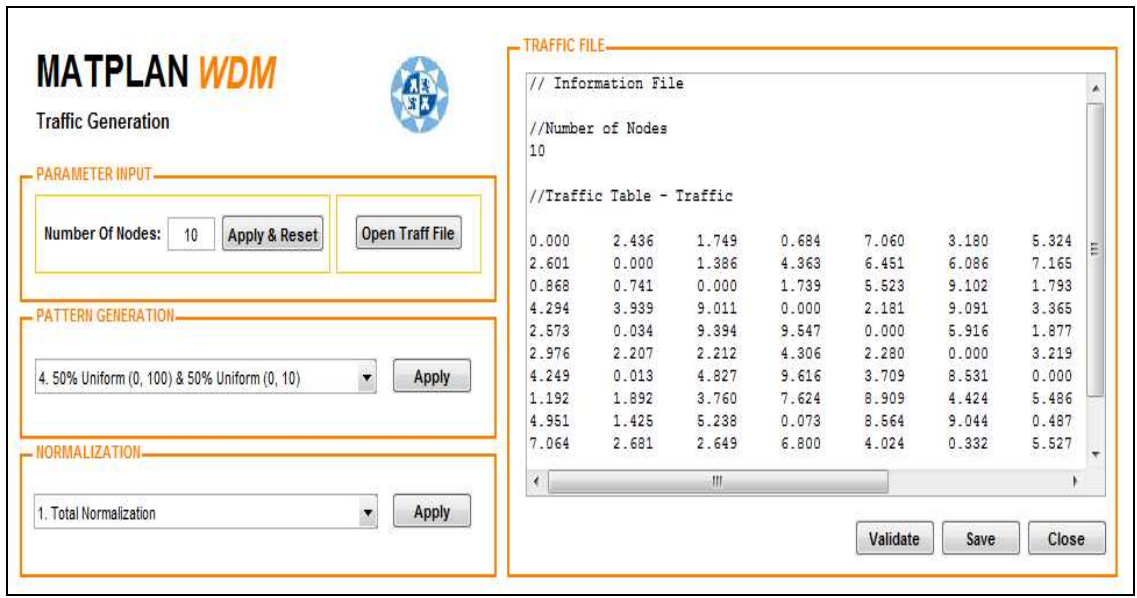

Figure 3.    Traffic generation for Topology 1

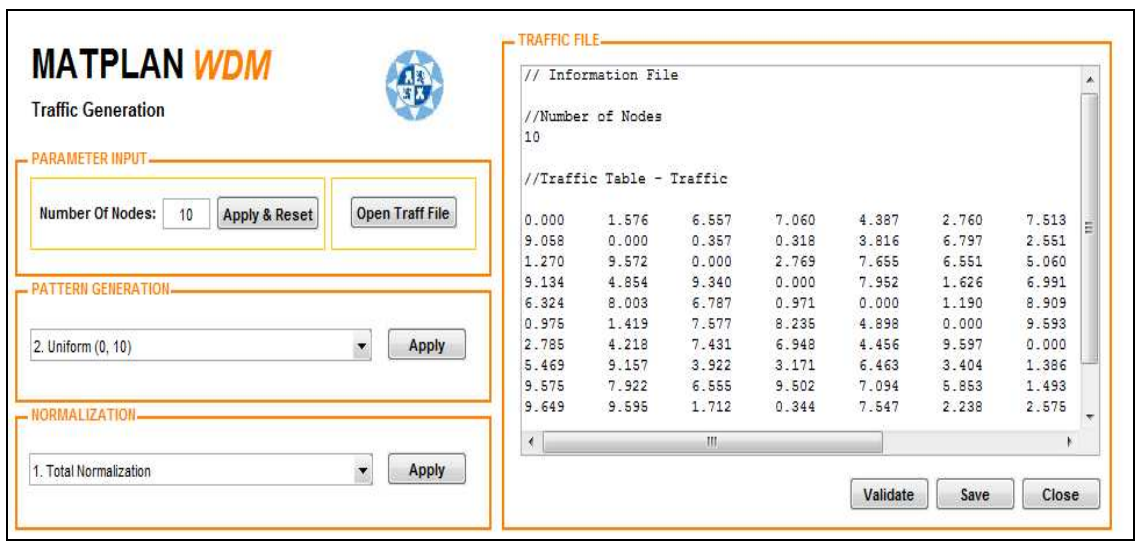

Figure 4.    Traffic generation for Topology 2

Optimization algorithms were applied for calculating a solution to the problem defined by the selected input parameters. These algorithms are implemented as MATLAB functions which follow a fixed signature, establishing the format of the input parameters, and the format of the output results. A broad range of heuristics were implemented - HLDA, MLDA, TILDA and RLDA. A new hybrid algorithm based on the following two algorithms were applied for routing of the multimedia traffic flows: (i) Capacitated Shortest Path First (CSPF) routing, which minimizes the distance of each flow linking the given source and destination nodes and satisfying capacity constraints; and (ii) Flow Deviation Algorithm (FDA) routing, which minimizes the network-wide average packet delay.

### 3. Simulation of New Algorithm and Results

The new hybrid algorithm attempts to accommodate the traffic flow in the existing virtual topology by minimizing the number of virtual hops. If it fails to accommodate the traffic flow, the algorithm tries to solve the problem by establishing one optical connection. It tries a direct optical connection from the source to destination node. If this fails, the set of nodes that are connected to the source node by a optical connection with enough capacity is calculated. Then, for each node in the set, the algorithm attempts to establish a optical connection from that node to the destination node. The node with the shortest delay is selected. If this fails, the algorithm computes the set of nodes that are connected to the destination node by a optical connection of enough capacity. For each node in the set, the algorithm attempts to establish a optical connection from the source node to that node. The node with the shortest delay is selected. If this fails, the flow is blocked. Each search of a optical connection from a source to a destination node is implemented by first obtaining the $k$-shortest paths in number of physical hops. For each possible route, a free wavelength is selected following the first-fit scheme. For a flow termination, the planning module removes the flow and checks if, as a consequence, any of the traversing optical connections becomes empty. The empty optical connections are then also terminated. Simulation results from the new hybrid algorithm are given in terms of the virtual topology and the routing table as shown in Fig. 5. The virtual topologies for Topology 1 and Topology 2 are given in Fig. 6 and Fig. 7 respectively. The complete routing performance is given in Fig. 8. Results show that the distance of each flow linking the given source and destination nodes was minimized given the capacity constraints and the total packet delay of the network was minimized.

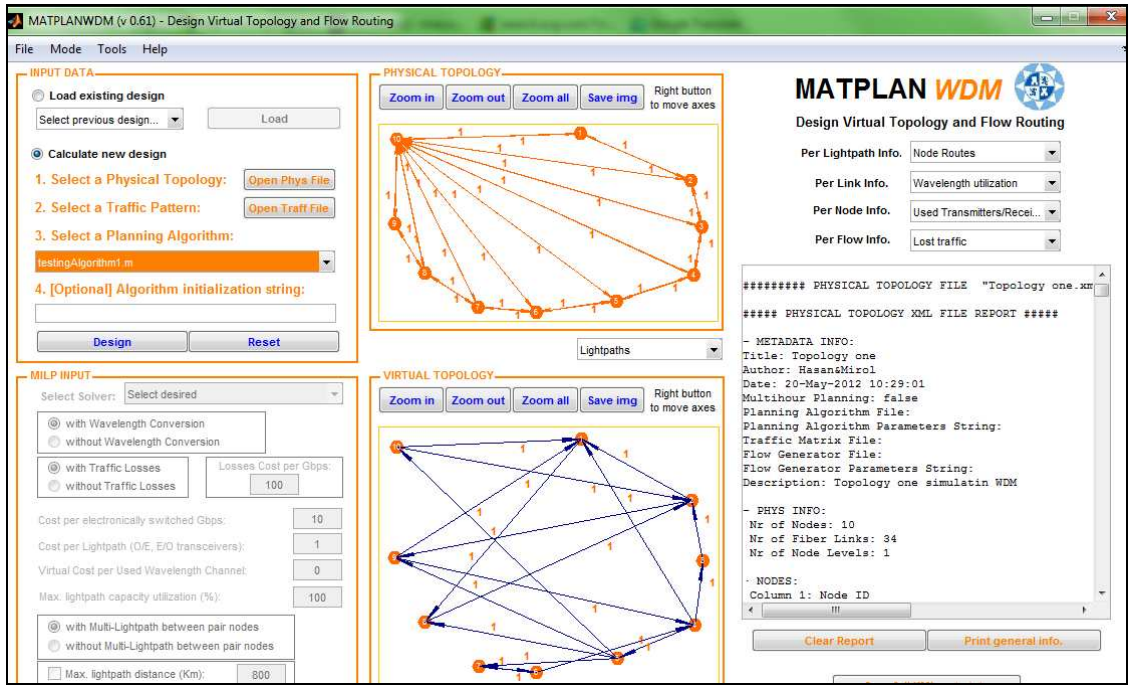

Figure 5.    Simulation results

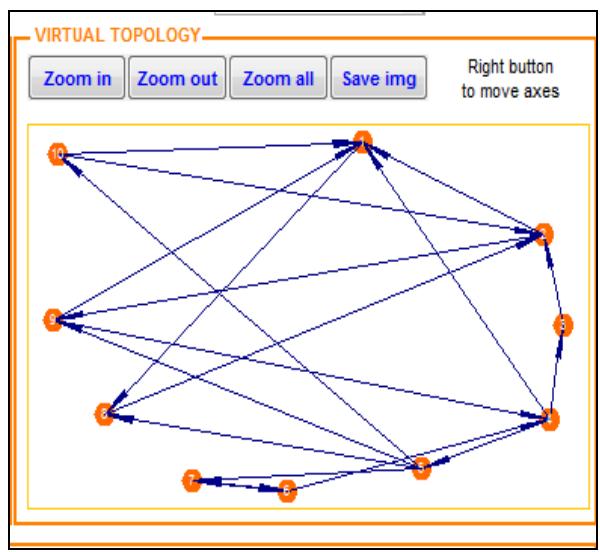

Figure 6.   Virtual topology for Topology 1

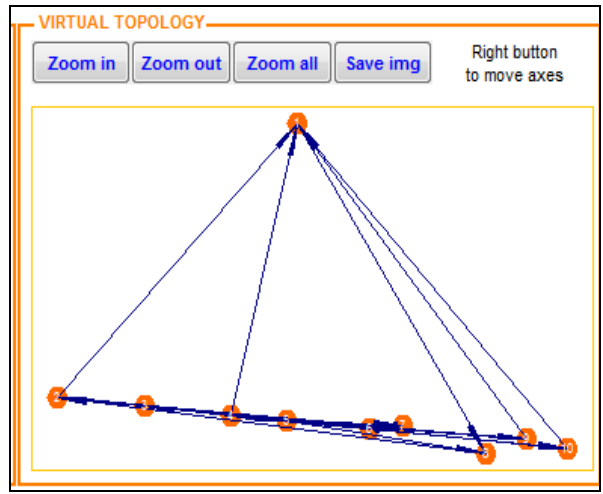

Figure 7. Virtual topology for Topology 2

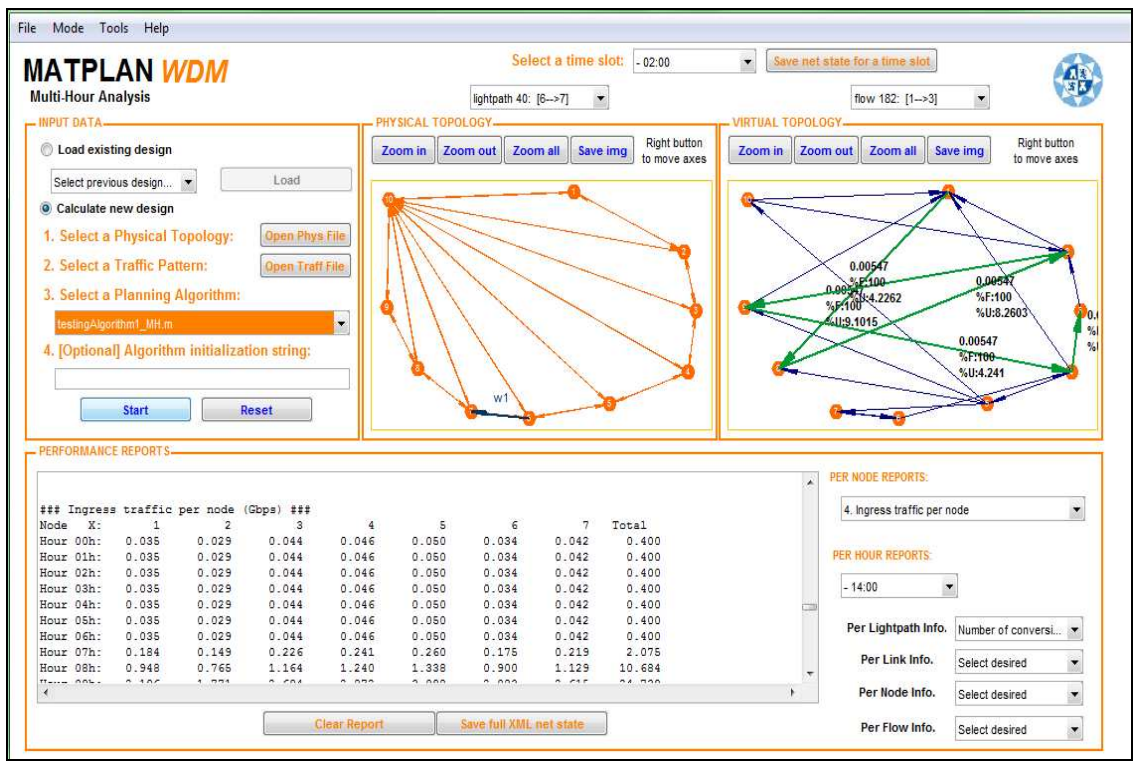

Figure 8.: Multi-Hour Multimedia Traffic Routing Performance

## 5. Conclusion

Multilayer wavelength division multiplexing (WDM) packet switching is a potential technology for increasing the bandwidth of optical networks. A multilayer WDM network with multimedia traffic was simulated using MatPlanWDM [13]. A new hybrid algorithm based on the Capacitated Shortest Path First (CSPF) and Flow Deviation Algorithm (FDA) was shown to successfully minimize the distance of each flow linking the given source and destination nodes, satisfies capacity constraints and minimizes the total packet delay of the network.